\documentclass[%
 amsmath,amssymb,
 aps,
 twocolumn
]{revtex4-2}

\usepackage{graphicx}
\usepackage{dcolumn}
\usepackage{bm}
\usepackage[colorlinks=true,citecolor=blue]{hyperref}
\usepackage{url}
\usepackage{subfigure}
\usepackage{listings}
\begin{document}

\title{A graphical user interface software for lattice QCD based on Python acceleration technology}

\author{Lin Gao}
\email{silvester\_gao@qq.com}

\affiliation{American Association for the Advancement of Science, Washington, DC 20005, USA}

\date{\today}

\begin{abstract}
A graphical user interface (GUI) software is provided for lattice QCD simulations, aimed at streamlining the process. The current version of the software employs the Metropolis algorithm with the Wilson gauge action. It is implemented in Python, utilizing Just-In-Time (JIT) compilation to enhance computational speed while preserving Python's simplicity and extensibility. Additionally, the program supports parallel computations to evaluate physical quantities at different inverse coupling \(\beta\) values, allowing users to specify the number of CPU cores. The software also enables the use of various initial conditions, as well as the specification of the save directory, file names, and background settings. Through this software, users can observe the configurations and behaviors of the plaquette under different \(\beta\) values.

\end{abstract}

\maketitle

\section{Introduction}
Currently, we recognize four fundamental interactions in nature: strong interaction, weak interaction, electromagnetic interaction, and gravitational interaction. Among these, gravitational and electromagnetic interactions are long-range interactions. Gravitational interaction plays a significant role in celestial dynamics, and the electromagnetic interaction can be observed in various forms at the macroscopic scale that our human eyes can directly see. In contrast, strong interaction is a short-range interaction, with effects typically observed at scales much smaller than those of electromagnetic interactions, usually at the femtometer level ($10^{-15} \, \text{m}$). This scale makes the observation of strong interactions particularly challenging. 

Moreover, strong interactions exhibit unique effects that are not directly applicable from studies of the macroscopic world, such as quarks carrying fractional electric charges, unlike macroscopic objects which can only possess integral charges. The mathematical framework describing strong interactions is also more complex than that for electromagnetic interactions; for example, electromagnetic interactions are described using abelian groups, while strong interactions require non-abelian groups.

These factors contribute to the relatively less comprehensive understanding of strong interactions compared to electromagnetic interactions, leaving many avenues for further research. Quantum Chromodynamics (QCD) is the theory dedicated to the study of strong interactions, quarks, gluons, and related phenomena. Perturbation theory is ineffective in the low-energy regime of QCD, necessitating the development of non-perturbative methods to address QCD problems. Lattice QCD represents one such non-perturbative approach that allows the study of QCD from first principles. In lattice QCD, we utilize formulations in Euclidean spacetime rather than Minkowski spacetime. Additionally, it is essential to discretize physical quantities defined in continuous spacetime, and the discrete gauge action used is the Wilson gauge action\cite{Wilson1974PRD,Gattringer2010}
\begin{equation}
{S_G}\left[ U \right] = \frac{\beta }{3}\mathop \sum \limits_{n \in {{\Lambda }}} \mathop \sum \limits_{\mu  < \nu} {\mathop{\rm Re}\nolimits} {\rm{Tr}}\left[ {1 - {U_{\mu \nu}}\left( n \right)} \right],
\end{equation}
where $\beta$ is the inverse coupling and $U_{\mu \nu}\left(n\right)$ is the plaquette
\begin{equation}
\begin{aligned}
U_{\mu \nu}(n) &= U_\mu(n) U_{\nu}(n+\hat{\mu})U_{-\mu}(n+\hat{\mu}+\hat{\nu}) U_{-\nu}(n+\hat{\nu}) \\
           &= U_\mu(n) U_{\nu}(n+\hat{\mu})U_{\mu}(n+\hat{\nu})^{\dagger}U_{\nu}(n)^{\dagger}.
\end{aligned}
\end{equation}
$U_\mu(n)$ is the link variable. A gauge invariant quantity about plaquette is as follows
\begin{equation}
{u_0} = {\left\langle {\frac{1}{3}Tr{U_{pl}}} \right\rangle ^{\frac{1}{4}}}.
\end{equation}
In this article, the configuration of the Wilson gauge field and $u_0$ will be calculated.

The Metropolis algorithm is used to handle the simulation of lattice QCD. In pure SU(3) lattice gauge theory, the conditional transition probability of a Markov process is\cite{Gattringer2010}
\begin{equation}
P\left(U_n=U^\prime\mid U_{n-1}=U\right)=T\left(U^\prime\mid U\right),
\end{equation}
where the configuration changes from $U$ to $U^\prime$. In many Monte Carlo algorithms, detailed balance condition is used 
\begin{equation}
T(U^\prime \mid U) P(U) = T(U \mid U^\prime) P(U^\prime),
\end{equation}
where $P(U)$ satisfies $P(U) \propto \exp(-S[U])$. Thus 
\begin{equation}
\frac{{T({U^\prime }\mid U)}}{{T(U\mid {U^\prime })}} = \frac{{P({U^\prime })}}{{P(U)}} = {e^{ - \Delta S}} \text{ with } \Delta S = S[U^\prime] - S[U].
\end{equation}
The conditional transition probability can be further written as the product of the priori selection probability $T_0(U^\prime \mid U)$ and the acceptance probability $T_A(U^\prime \mid U)$. Therefore, when the priority selection probability has symmetry
\begin{equation}
T_0(U \mid U^\prime) = T_0(U^\prime \mid U),
\end{equation}
we can obtain
\begin{equation}
\frac{{{T_A}({U^\prime }\mid U)}}{{{T_A}(U\mid {U^\prime })}} = {e^{ - \Delta S}}.
\end{equation}
In the Metropolis algorithm, the acceptance probability $T_A(U^\prime \mid U)$ can be simplified to\cite{metropolis1953equation}
\begin{equation}
T_A(U^\prime \mid U) = \min \left(1, \exp(-\Delta S) \right) ,
\end{equation}
where $\Delta S = S[U^\prime] - S[U]$.

\section{Improvement of Computational Speed}

The enhancement of computational speed arises from both hardware and software improvements. 

From a hardware perspective, a combination of CPUs and GPUs can be employed to accelerate computations. Intel's Gordon Moore famously proposed Moore's Law\cite{moore1965,moore2006progress}, which states that the number of transistors (or MOSFETs) on a computer chip doubles approximately every two years (or 18 months in some versions), leading to a corresponding doubling of microprocessor performance every 18 months. This law dominated chip development for an extended period\cite{Leiserson2020}; however, Intel's production of 14-nanometer chips in 2014 was followed by a delay in the introduction of its 10-nanometer process until 2019. While some companies have claimed successful research into 7-nanometer or even smaller sizes, many current technology nodes have become equivalent dimensions rather than actual channel lengths of MOSFETs. From a physical standpoint, as MOSFET sizes continue to shrink, issues such as increased tunneling currents and decreased effective carrier mobility in the channel become more prevalent. Additionally, the radius of a silicon atom is approximately 111 picometers. These aspects indicate that the size of individual MOSFET is nearing a physical limit. Hardware structure also plays a significant role; for instance, when CPUs and GPUs are manufactured using the same semiconductor technology, GPUs generally outperform CPUs in matrix parallel computations due to their architectural advantages.

On the software side, while programs can be written more concisely, this may lead to decreased performance efficiency. Therefore, it is essential to find ways to enhance the calculation speed of software. One study demonstrated that researchers were able to multiply two \(4096 \times 4096\) matrices by parallelizing the code to run across all 18 processing cores, optimizing the memory hierarchy of the processor, and utilizing Intel's Advanced Vector Extensions (AVX) instructions. The optimized code completed the computation in just 0.41 seconds, whereas Python 2 required 7 hours and Python 3 required 9 hours\cite{Leiserson2020}. This underscores the importance of software-hardware synergy in improving computational speed.

Python allows for concise and highly extensible programming. However, Python code typically exhibits slower execution speeds, necessitating acceleration techniques to enhance computational efficiency. Just-In-Time (JIT) Compilation is a method that improves program execution efficiency by dynamically compiling bytecode or other intermediate code into machine code during program execution\cite{lam2015numba}. This allows the program to run directly on machine code, thereby reducing the overhead associated with interpretation.

In this study, Numba's Just-In-Time (JIT) compilation is utilized to accelerate Python computations \cite{numpy2020array, lam2015numba}. For detailed results on the speed improvement, see the Results and Discussion section.

\section{Main Features of the Software}

\subsection{Graphical User Interface (GUI)}
The GUI of the program is developed using Tkinter, providing an intuitive interface for parameter setup and simulation execution. The layout is modernized with custom fonts and colors, enhancing the user experience. Users can set lattice size, \(\beta\), iterations, CPU core numbers, and initial scheme.

\subsection{Custom Background Images}
The GUI allows for custom background images to enhance visual appeal. Users can personalize the background by replacing the \texttt{background.jpg} file used in the GUI.

\subsection{Initialization Schemes and Boundary Conditions}
The program supports two types of initial lattice schemes (hot start or cold start), enabling users to explore various initial conditions. Additionally, periodic boundary conditions are employed.

\subsection{Parallel Processing}
Users can process simulations for multiple \(\beta\) values in parallel, specifying the number of CPU cores to optimize the utilization of multi-core systems. The program leverages Python's \texttt{multiprocessing} module to efficiently handle simulations across different \(\beta\) values, executing each simulation in independent processes that update results separately.

Currently, the version of the program does not implement Numba parallelization for matrix multiplication within one configuration. This decision is based on the observation that such parallelization involves frequent thread activation and deactivation, which can be time-consuming and less efficient compared to parallel simulations across different \(\beta\) values. However, if computations are limited to a single \(\beta\) or if an exceptionally high number of CPU cores are available, internal parallel computations may still be considered.

\subsection{Visualization and Data Saving}
Upon completion of the simulation, the program saves the final lattice configuration and \(u_0\) data in \texttt{.npy} format, automatically storing them in the specified directory. The \texttt{.npy} format is a file format used by NumPy \cite{numpy2020array}. Graphs illustrating the variation of \(u_0\) with iteration counts are generated for each \(\beta\), assisting users in understanding the temporal evolution of the system. The visualization utilizes Matplotlib for saving images\cite{hunter2007matplotlib}.

\section{User Instructions}

\begin{figure}[htb]
    \centering
    \includegraphics[width=0.8\linewidth]{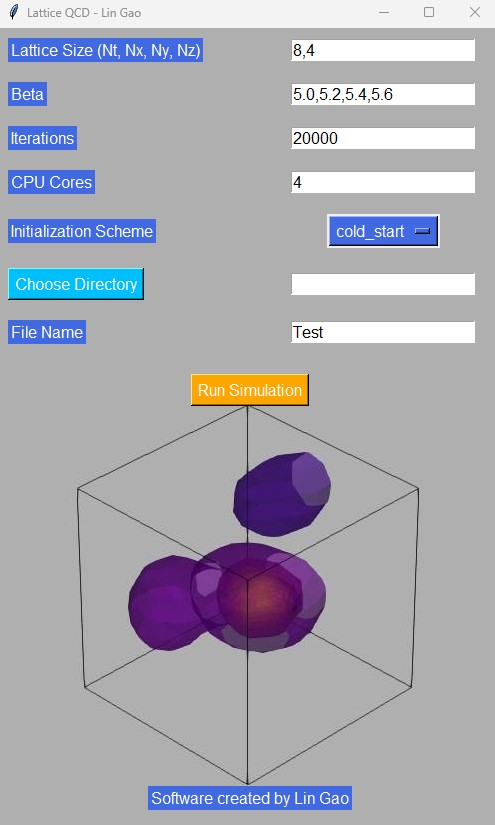}
    \caption{An example of parameters.}
    \label{LQCD_GUI}
\end{figure}

Refer to Fig.~\ref{LQCD_GUI} for an example of input parameters. In the GUI, set the required parameters (\textbf{lattice size}, \(\beta\), iteration counts, CPU core numbers, and initialization scheme).

1. \textbf{Lattice Size}: The lattice size supports two input methods. When \(N_x = N_y = N_z\), only \(N_t\) and \(N_x\) need to be specified. The program will automatically interpret these as \(N_t\), \(N_x\), \(N_y\), and \(N_z\). In this example, the input ``8,4'' indicates \(N_t = 8\) and \(N_x = N_y = N_z = 4\). If \(N_x\), \(N_y\), and \(N_z\) are not equal, all four parameters must be provided, separated by commas.

2. \textbf{\(\beta\)}: Multiple values for \(\beta\) can be input simultaneously to facilitate parallel computations. Different values should be separated by commas.

3. \textbf{Iteration Counts}: One complete update of all lattice points is called one sweep. Furthermore, it is computationally economic to repeat the updating step 10 times for the visited variable, since the computation of the sum of staples is costly\cite{Gattringer2010}. In this context, one iteration corresponds to 10 sweeps.

4. \textbf{Initialization Scheme}: The default option is a cold start; however, users can also select a hot start.

5. \textbf{File Saving Options}: Users can specify the save directory and file name. In this example, if the file name is designated as ``Test", all file names of generated configurations and plaquette-related files will begin with ``Test".

After completing these parameter settings, press the ``Run Simulation'' button to initiate the simulation.

\section{Results and Discussion}
This paper presents a lattice QCD simulation program based on the Metropolis algorithm, utilizing a GUI to facilitate intuitive user input, thereby visualizing and simplifying the simulation process. This implementation demonstrates the program's advantages in user-friendliness. In this section, a discussion will focus on the computational speed of the JIT-optimized Python 3. Additionally, further testing will be conducted on the data generated by this GUI software.

\subsection{computational speed}
Both C++ and JIT-optimized Python 3 programs were developed for multiplying two \(SIZE \times SIZE\) matrices, measuring the computation time. The relevant codes are provided in Appendix A and Appendix B. As shown in Fig. \ref{fig_time_cost}, the computation time for matrix multiplication increases as the matrix size (\(SIZE\)) grows from 200 to 2000 for both methods. 
\begin{figure}[htb]
    \centering
    \subfigure[The time cost for multiplying two \(SIZE \times SIZE\) matrices using the C++ program]{
        \includegraphics[width=0.5\textwidth]{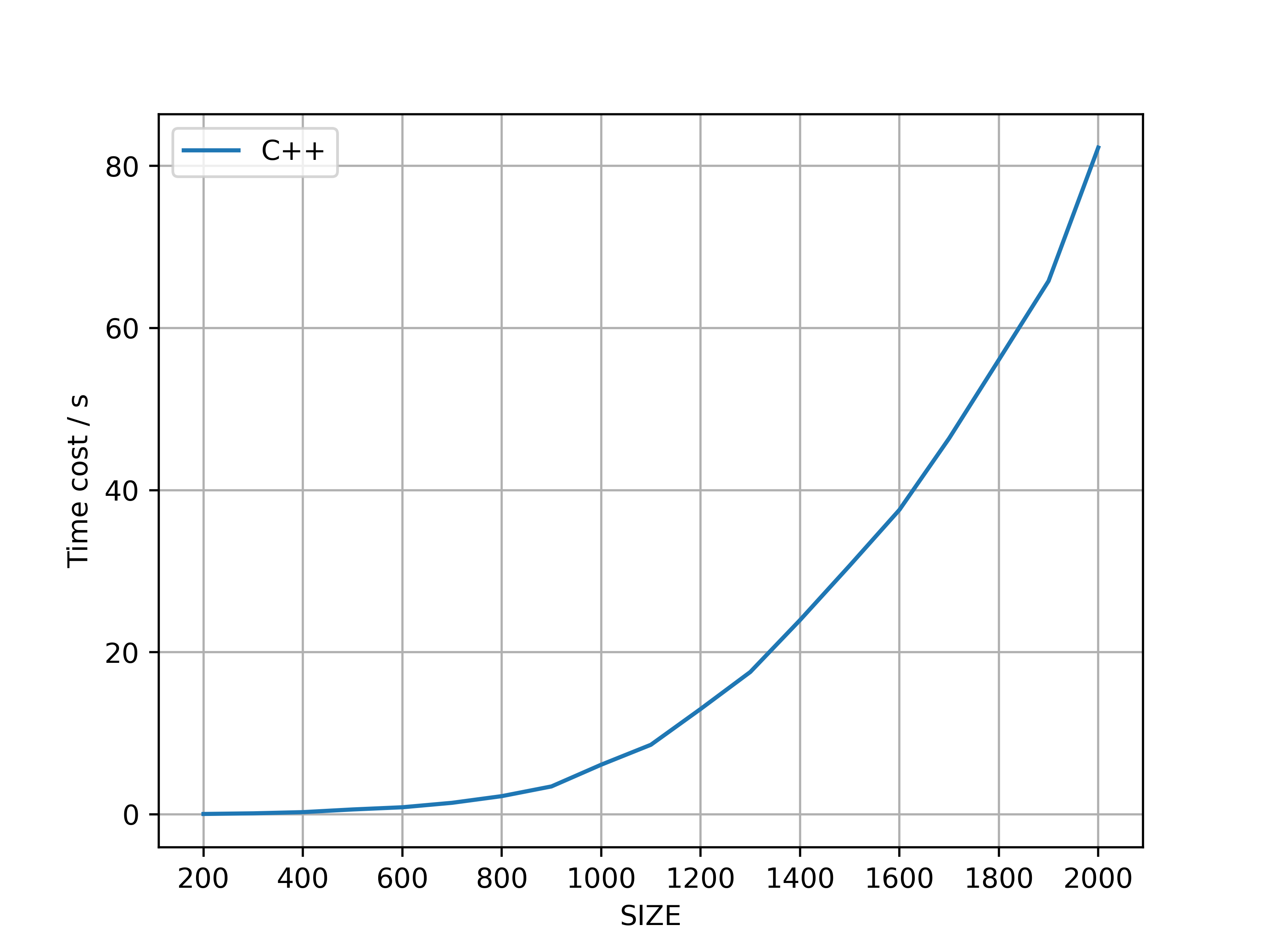}
    }
    \hspace{0.05\textwidth} 
    \subfigure[The time cost for multiplying two \(SIZE \times SIZE\) matrices using the JIT-optimized Python 3 program]{
        \includegraphics[width=0.5\textwidth]{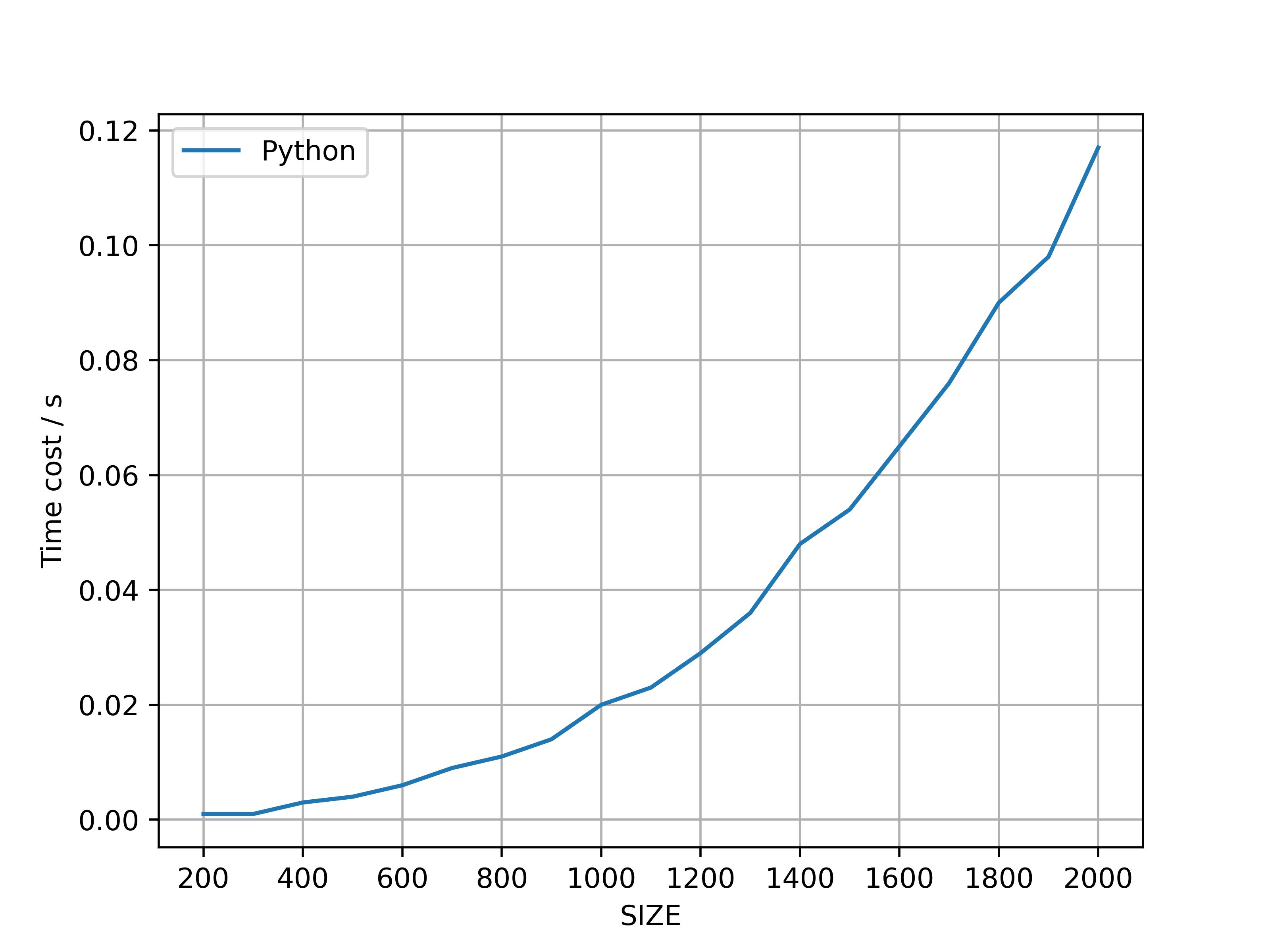}
    }
    \caption{The time cost for multiplying two $SIZE \times SIZE$ matrices is evaluated, with $SIZE$ incrementally increasing from 200 to 2000 in intervals of 100. For each matrix size, matrix multiplication is implemented using both C++ and JIT-optimized Python3.}
    \label{fig_time_cost}
\end{figure}

The time complexity of matrix multiplication is typically \(O(SIZE^3)\). Therefore, for two \(SIZE \times SIZE\) matrices, the time required for multiplication is approximately proportional to \(SIZE^3\). Consequently, when the matrix size doubles (i.e., \(SIZE\) becomes \(2 \times SIZE\)), the computation time increases by a factor of eight. In this test, using C++ as an example, the following time costs were observed for smaller matrix sizes: \(SIZE=200\) resulted in \(0.033 s\), \(SIZE=400\) in \(0.259 s\), and \(SIZE=800\) in \(2.232 s\), which aligns with the expected eightfold growth. However, for larger matrices, such as \(SIZE=1600\), the time cost escalated to \(37.565 s\), leading to the observation that \(37.565/2.232 \gg 8\). 

Several factors contribute to this phenomenon. Matrix multiplication involves extensive data reading and writing. When the matrix size exceeds the CPU's cache capacity, memory access becomes more frequent, leading to increased computation time. Furthermore, matrix multiplication requires significant data transfer, particularly for large matrices, which can create a memory bandwidth bottleneck. Additionally, matrix rows and columns are stored linearly in memory; accessing matrices—especially in a column-major order—can result in poor cache utilization, negatively impacting performance. Moreover, the standard triple-loop algorithm is inefficient for large-scale matrices. Although more efficient algorithms exist, their complexities may not be straightforward, and their implementation can be complex. Consequently, due to the aforementioned factors, computation time often increases significantly in practical applications.

The results indicate that the JIT-optimized Python3 program for multiplying two \(2000 \times 2000\) double-precision matrices took \(0.117s\), while the C++ implementation took \(82.268s\). It is evident that, thanks to the optimizations in matrix multiplication provided by NumPy and Numba, the JIT-optimized Python3 implementation is faster than the C++ triple-loop algorithm. Moreover, the program is significantly more concise, which is advantageous when writing larger programs.

\subsection{Configuration Testing}
The software was tested with the parameters Lattice Size (Nt, Nx, Ny, Nz) set to (8, 4, 4, 4), $\beta = 5.0, 5.2, 5.4, 5.6$, iteration = 2000, and the number of CPU cores set to 4. The key results and discussions are presented below.

The generated configurations consist of some SU(3) matrices, specifically with dimensions of ${N_t} \times {N_x} \times {N_y} \times {N_z} \times 4 \times 3 \times 3$, where the last $3 \times 3$ indicates the size of the SU(3) matrices. This dimension is derived under the consideration of complex values. If only real values are considered, the overall size becomes ${N_t} \times {N_x} \times {N_y} \times {N_z} \times 4 \times 3 \times 3 \times 2$.

A thorough examination of all the $3 \times 3$ complex matrices in the configuration revealed that they are indeed SU(3) matrices.

\subsection{Plaquette Testing}

\begin{figure}[h]
    \centering
    \includegraphics[width=1.0\linewidth]{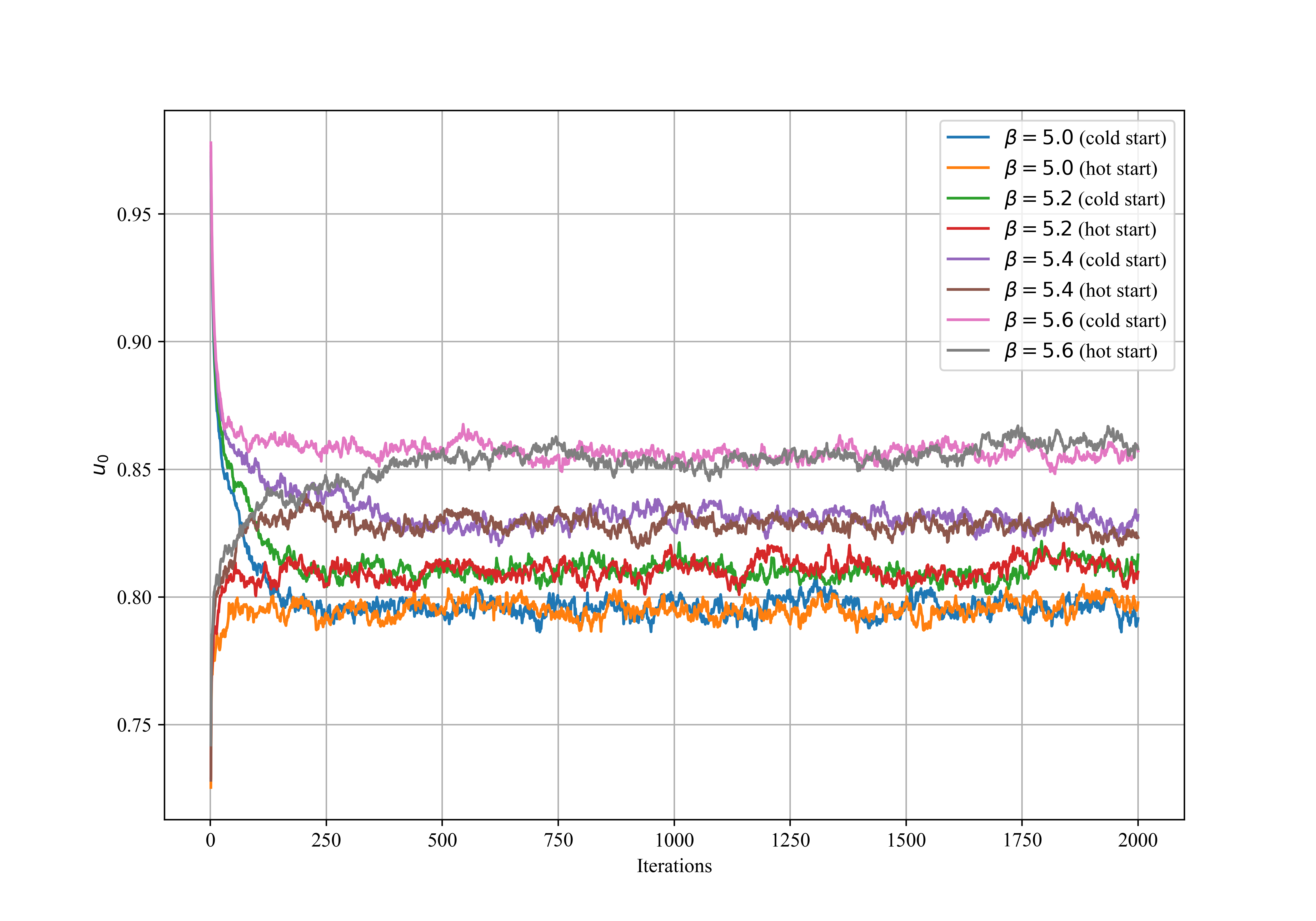}
    \caption{ The evolution of $u_0$ with respect to the iteration under different conditions is illustrated. This figure highlights how $u_0$ varies as the iterations progress, demonstrating the impact of varying conditions on the system's behavior. Such observations are crucial for understanding the dynamics of the system and validating the simulation results.
}
    \label{plaq_beta}
\end{figure}

As illustrated in Fig.~\ref{plaq_beta}, after 1000 iterations, the curves for the same $\beta$ starting from different initial conditions converge. This convergence is observed for all tested $\beta$ values, indicating that the system reaches equilibrium after 1000 iterations.

The software allows for intuitive input of parameters such as lattice size and iterations, executing the Metropolis algorithm's update process in parallel and achieving system equilibrium. This demonstrates the software's effectiveness in lattice QCD simulations. The actual results show that with adjustments to $\beta$ and iteration counts, the physical quantity $u_0$ exhibits reasonable variations that align with physical expectations.

\section{conclusion}
This paper presents a graphical user interface (GUI) software for lattice QCD based on Python acceleration techniques, achieving a complete workflow from input parameters to result output. This approach offers a new perspective for studying numerical simulations of lattice QCD, facilitating wider user adoption and understanding through an intuitive GUI design. The simulation software provides an easy-to-use platform that combines parallel computing and interactive visualization, assisting users in performing lattice QCD calculations. The main conclusions are as follows:

\textbf{User Friendliness and Experience.}  
The GUI facilitates parameter input and result saving, allowing users to conduct physical simulation experiments without writing code. This interface design is particularly appealing to students, users outside computational physics, and researchers in experimental physics who may not be adept at programming. Additionally, the software incorporates customizable background images, enabling users to modify the background according to their preferences. The background images enhance the visual feedback, making the simulation experience more vivid and intuitive, thereby highlighting the importance of visual elements in improving interface friendliness. Furthermore, the software design emphasizes flexibility, permitting users to freely adjust simulation parameters according to specific research needs. Users can input lattice size, set different ranges for \(\beta\), and choose initial conditions, accommodating a variety of experimental contexts.

\textbf{Python Acceleration Techniques.}  
Python offers simplicity and high scalability in programming. However, traditional Python programs are significantly slower than C/C++ for numerical calculations. This study utilizes Just-In-Time (JIT) compilation techniques to accelerate Python computations, preserving Python's simplicity while enhancing its computational speed. In the example in this article, the JIT-optimized Python program for matrix multiplication even outperforms the traditional C++ triple-loop algorithm.

\textbf{Parallel Computing.}  
The software's parallel computing capability significantly improves simulation efficiency. By fully leveraging the computational power of multi-core processors, users can obtain simulation results under different \(\beta\) values in a shorter time. This efficiency not only conserves computational resources but also enables researchers to conduct larger-scale experiments, exploring a broader parameter space and thereby advancing physical research.

\textbf{Diverse Data Output and Research Applicability.}  
The program outputs include the final lattice configurations and the \(u_0\) data as a function of iteration, saved in $.npy$ format. Additionally, graphs depicting the variation of \(u_0\) with iteration will be generated for each \(\beta\). These outputs provide researchers with diverse options for post-processing and data analysis. The output images illustrate the evolution of \(u_0\) under different \(\beta\) values and initialization schemes, visually reflecting the impact of model parameters on the system state. The $.npy$ data files facilitate loading into other computational environments for further analysis and data mining, thereby providing researchers with convenient data management and analysis options.

\textbf{Future Work.}  
Overall, the functionality of this program effectively implements stable simulations using the Metropolis algorithm and demonstrates excellence in numerical simulation and user interface friendliness. The analysis and discussion of the results indicate that this method lays a solid foundation for further model expansion and performance optimization. Due to Python's high scalability, this software is easily extendable to incorporate more functionalities in future versions, including additional physical quantities and alternative actions. Currently, the software only supports simple periodic boundary conditions, and future versions may consider other boundary conditions. Furthermore, machine learning techniques are increasingly influencing lattice QCD research, and integration of machine learning content will be considered in subsequent versions\cite{gao2024}. Lastly, further optimization of the parallelization aspect will include the potential addition of GPU acceleration.

\section{Software Acquisition}
The link to this GUI software is as follows:
\url{https://drive.google.com/file/d/1f0XgoQmge_hFssSsSxWd5C0s2aYYc_qF/view?usp=sharing}.

You may need the following.
\begin{itemize}
  \item Python 3.x
  \item Some packages: \texttt{numpy}, \texttt{numba}, \texttt{matplotlib}
\end{itemize}

\section*{Acknowledgments}  This research utilized ChatGPT to polish the paper. I extend my gratitude to the contributors of GPT for their valuable contributions\cite{openai2023gpt4,chatgptreasoning}.

\appendix
\begin{widetext}
\section{Python program for matrix multiplication}
\begin{lstlisting}[language=Python, caption=Python code]
import numpy as np
import time
from numba import njit,set_num_threads

set_num_threads(1)


# Matrix multiplication function
@njit
def matrix_multiply(A, B):
    result = np.dot(A, B)
    return result


for SIZE in range(100,2100,100):

    # The start time
    start_time = time.time()

    # Generate two random matrices
    matrix1 = np.random.rand(SIZE, SIZE)
    matrix2 = np.random.rand(SIZE, SIZE)

    # Matrix multiplication
    result = matrix_multiply(matrix1, matrix2)

    # The end time and the time taken
    end_time = time.time()
    print(f"{SIZE} {end_time - start_time:.3f}")

\end{lstlisting}

\section{C++ program for matrix multiplication}
\begin{lstlisting}[language=C++, caption=C++ code]
#include <iostream>
#include <cstdlib>
#include <ctime>


// Matrix multiplication function
void multiplyMatrices(double** matrix1, double** matrix2, double** result,int SIZE) {
    for (int i = 0; i < SIZE; ++i) {
        for (int j = 0; j < SIZE; ++j) {
            result[i][j] = 0;
            for (int k = 0; k < SIZE; ++k) {
                result[i][j] += matrix1[i][k] * matrix2[k][j];
            }
        }
    }
}

int main() {

    clock_t start_time,end_time;

    for (int SIZE = 100; SIZE < 2100; SIZE += 100) {

        //  The start time
        start_time = clock();


        // Create random matrices matrix1 and matrix2
        double **matrix1 = new double *[SIZE];
        double **matrix2 = new double *[SIZE];
        double **result = new double *[SIZE];

        for (int i = 0; i < SIZE; ++i) {
            matrix1[i] = new double[SIZE];
            matrix2[i] = new double[SIZE];
            result[i] = new double[SIZE];
        }


        std::srand(static_cast<unsigned int>(std::time(nullptr)));

        // Fill matrices matrix1 and matrix2 with random numbers
        for (int i = 0; i < SIZE; ++i) {
            for (int j = 0; j < SIZE; ++j) {
                matrix1[i][j] = static_cast<double>(std::rand()) / RAND_MAX;
                matrix2[i][j] = static_cast<double>(std::rand()) / RAND_MAX;
            }
        }

        // Multiply the matrices
        multiplyMatrices(matrix1, matrix2, result, SIZE);


        for (int i = 0; i < SIZE; ++i) {
            delete[] matrix1[i];
            delete[] matrix2[i];
            delete[] result[i];
        }
        delete[] matrix1;
        delete[] matrix2;
        delete[] result;

        //The end time and the time taken
        end_time = clock();
        std::cout<< SIZE<<" "<< (double) (end_time-start_time) / CLOCKS_PER_SEC<< std::endl;

    }

    return 0;
}
\end{lstlisting}
\end{widetext}

\bibliographystyle{apsrev4-2}
\bibliography{apssamp}

\end{document}